\RequirePackage{fix-cm}
\documentclass{article}       % onecolumn (second format)

\usepackage{graphicx}
\usepackage{marvosym}
\usepackage{algorithm}
\usepackage{algorithmic}
\usepackage{amsmath}
\usepackage{bm}
\usepackage[usenames,dvipsnames]{color}
\usepackage{subfig}
\usepackage{amssymb}
\usepackage{wasysym}
\usepackage{multirow}
\usepackage{authblk}

\newcommand{\vect}[1]{\mathbf{#1}}

\newcommand{\idt}{$\quad$ }

\newcommand{\kHz}{\text{kHz}}

\begin{document}
{\huge{This paper was accepted for publication in Circuits, Systems \& Signal Processing journal. A copyright may be transferred without notice.}}

\newpage
% ----- ----- ----- ------ AUTHORS AND INSTITUTES ----- ----- ----- -----
\author{Jacek Pierzchlewski, Thomas Arildsen}
\affil{Signal and Information Processing, \\ Department of Electronic Systems, Aalborg University, \\ 
Fredrik Bajers Vej 7, DK-9220 Aalborg, Denmark \\
jap@es.aau.dk, tha@es.aau.dk}

% ----- ----- ----- ------ RECEIVED / ACCEPTED IS HERE  ----- ----- ----- -----

% ----- ----- ----- ------ ----- TITLE ----- ----- ----- ----- -----
\title{Generation and Analysis of Constrained Random Sampling Patterns}
\maketitle

% ----- ----- ----- ------ ABSTRACT STARTS HERE ----- ----- ----- -----
\begin{abstract}
Random sampling is a technique for signal acquisition which is gaining popularity in
practical signal processing systems.
Nowadays, event-driven analog-to-digital converters make random sampling
feasible in practical applications.
A process of random sampling is defined by a sampling pattern,
which indicates signal sampling points in time.
Practical random sampling patterns are constrained by ADC characteristics
and application requirements.
In this paper we introduce statistical methods which evaluate random
sampling pattern generators with emphasis on practical applications.
Furthermore, we propose a new random pattern generator which copes
with strict practical limitations imposed on patterns,
with possibly minimal loss in randomness of sampling.
The proposed generator is compared with existing sampling pattern generators
using the introduced statistical methods.
It is shown that the proposed algorithm generates
random sampling patterns dedicated for event-driven-ADCs better than existed
sampling pattern generators.
Finally, implementation issues of random sampling patterns are discussed.

% ----- ----- ----- ------ ----- KEYWORDS ----- ----- ----- ----- -----
{\bf{keywords:}} Analog-digital conversion, Compressed sensing, Digital circuits, Random sequences, Signal sampling

\end{abstract}

%-------------------------------------------------------------------------
%  ----- ----- ----- INTRODUCTION
%-------------------------------------------------------------------------
\section{Introduction}
In many of today's signal processing systems there is a need for random signal sampling.
The idea of random signal sampling dates back to early years
of the study on signal processing \cite{Sha01}.
Signal reconstruction methods for this kind of sampling were studied \cite{Feichtinger95},
there are practical implementations of signal acquisition systems
which employ random nonuniform sampling \cite{Hom09,Hui01,Wakin12}.
Recently, this method of sampling has received more attention hence to
a relatively new field of signal acquisition known as
compressed sensing \cite{Cand01,Las01}.
It was shown that in many compressed sensing applications
the random sampling is a correct choice for signal acquisition \cite{Bar01}.
The random sampling gives a possibility to sample
below Nyquist rate, which lowers the power dissipation
and reduces the number of samples to be processed.
A process of random sampling is defined by a sampling pattern,
which indicates signal sampling points in time.
Generation and analysis of random sampling patterns which
are dedicated to be implemented in analog-to-digital converters
is a subject of this work.

In practice, sampling according to a given sampling pattern
is realized with analog-to-digital converters \cite{An01,Le05}.
Currently, there are available event-driven analog-to-digital converters, which are able to realize
random sampling \cite{Hui01,Xil01}.
These converters have certain practical constraints coming from implementation
issues, which consequently puts implementation-related constraints on sampling patterns.
These constraints concern minimum and maximum time intervals between
adjacent sampling points,
e.g. Wakin et. al. in their work \cite{Wakin12} used a random nonuniform sampling pattern with minimum and maximum intervals between
adjacent sampling points.
Furthermore, there are application-related constraints
which concern stable average sampling frequency of sampling patterns,
equal probability of occurrence of possible sampling points,
and uniqueness of generated patterns.

The problem which this work solves is composed of two parts.
Firstly, how to evaluate different sampling pattern generators with emphasis on practical applications?
The early work on estimation of random nonuniform sampling patterns was done by Marvasti \cite{Marvasti89}.
Wakin et. al. \cite{Wakin12} looked for a sampling pattern with the best (most equal) histogram of inter-sample spacing.
Gilbert et. al \cite{Gil01} proposed to choose a random sampling pattern based on permutations.
To the best of our knowledge, 
there is no scientific work published which concerns multiparameter statistical analysis of random sampling patterns.
Due to the constantly increasing available computational power
it has become possible to analyze random pattern generators statistically
within a reasonable time frame.
Statistical parameters which assess random
sampling pattern generators with respect to the constraints described above are described in this paper.

The second problem discussed in this paper is how to 
construct a random sampling pattern generator which generates patterns with a given number of
sampling points, and given intervals between sampling points, with possibly minimum loss in randomness?
The well known random sampling pattern generators are Additive Random Sampling (ARS)
and Jittered Sampling (JS) \cite{Marvasti93,Woj01}.
However, these sampling pattern generators do not take into account
the mentioned implementation constraints, which is an obstacle in practical applications.
There have been some attempts to generate more practical sampling patterns.
Lin and Vaidyanathan \cite{Lin01} discussed periodically nonuniform sampling patterns 
which are generated by employing two uniform patterns.
Bilinskis et al. in \cite{Bilin01} introduced a concept of correlated additive random sampling, 
which is a modification of the ARS.
Papenfu\ss{} et al. in \cite{Pap01} proposed another modification of the ARS process, 
which was supposed to optimally utilize the ADC.
Ben-Romdhane et al. \cite{Rom08} discussed a hardware implementation of a nonuniform pseudorandom clock generator.
Unfortunately, none of the proposed sampling pattern generators are designed to address all the implementation constrains.
This paper proposes a sampling pattern generator which is able to produce constrained random sampling patterns
dedicated for use in practical acquisition systems.
The generator is compared with existing solutions using the proposed statistical parameters.
Implementation issues of this generator are discussed.

The paper is organized as follows.
The problem of random sampling patterns generation is identified in Section
\ref{sec:problem}.
Statistical parameters for random pattern generators are proposed in Section
\ref{sec:parameters}.
A new random sampling pattern generator for patterns to be used in practical
applications is proposed in Section \ref{sec:generators}.
The proposed generator is compared with existing generators in Section
\ref{sec:taa}.
Some of the implementation issues of random sampling patterns are discussed in
Section \ref{sec:implem}.
Conclusions close the paper in Section \ref{sec:conclusions}.
The paper follows the reproducible research paradigm \cite{Vand01}, therefore
all of the code associated with the paper is available online \cite{Jap01}.

%-------------------------------------------------------------------------
%  ----- ----- ----- PROBLEM FORMULATION
%-------------------------------------------------------------------------
\section{Problem formulation}
\label{sec:problem}

%2.1
\subsection{Random sampling patterns}
\label{subsec:patterns}

          This paper focuses on generation and analysis of random sampling patterns.
          The purpose of this Section is to formally define a sampling pattern and its parameters,
          and to discuss requirements for sampling patterns and sampling pattern generators.
        A sampling pattern $\mathbb{T}$ is an ordered set (sequence) with $K_{\text{s}}$ fixed sampling time points:
        \begin{equation}
                \mathbb{T} = \{ t_{1}, t_{2},\ldots,t_{K_{\text{s}}} \} \quad
        \end{equation}
        where the sampling time points $t_k$ are real numbers ($t_k \in \mathbb{R} , \; k = \{1, 2, \ldots, K_{\text{s}} \}$).
        Elements of such a set $\mathbb{T}$ must increase monotonically:
        \begin{equation}
        \label{eq:mono}
                t_{1} < t_{2} < \ldots < t_{K_{\text{s}}}
        \end{equation}
        Time length $\tau$ of a sampling pattern is equal to the time length of a signal or a signal segment on which the sampling pattern is applied.
        The time length $\tau$ may be higher than the last time point in a pattern: $\tau \geq t_{K_{\text{s}}}$.

        Any sampling point $t_{k} \in \mathbb{T}$ is a multiple of a sampling grid period $T_{\text{g}}$:
        \begin{equation}
                t_k = k{T_{\text{g}}}, \quad k \in \mathbb{N}^{\star}
        \end{equation}
        where $\mathbb{N}^{\star}$ is the set of natural numbers without zero.
        The sampling grid is a set:
        \begin{equation}
                \label{eqalg:grid}
                \mathbb{G} = \{ T_{\text{g}},\, 2T_{\text{g}},\, \ldots ,{K_{\text{g}}}{T_{\text{g}}} \}, \quad K_{\text{g}} = \left \lfloor \frac{\tau}{T_{\text{g}}} \right \rfloor
        \end{equation}
        where $K_{\text{g}}$ is the number of sampling grid points in a sampling pattern,
        and $\left \lfloor \cdot \right \rfloor$ signifies the floor function, which returns the largest integer lower or equal to the function argument.
        It can be stated that a pattern $\mathbb{T}$ is a subset of a grid set $\mathbb{G}$ ($\mathbb{T} \subset \mathbb{G})$.
        The sampling grid period $ T_{\text{g}}$ describes the resolution of the sampling process.
        In practice, the lowest possible sampling grid depends on the performance of the used ADC,
        its control circuitry, and the clock jitter conditions \cite{Hui01,An01,Le05}.
          A sampling pattern may be represented as indices of sampling grid:
          \begin{equation}
          \label{eq:grid_representation}
            \mathbb{T}'=\{t'_{1}, t'_{2}, \ldots, t'_{K_{\text{s}}} \}, \quad t'_{k} = \frac{t_{k}}{K_{\text{g}}}
          \end{equation}

        Let us define a set $\mathbb{D}$ which contains $K_{\text{s}}-1$ intervals between the sampling points:
        \begin{equation}
        \label{eq:distset}
          \mathbb{D} = \{ d_{1}, d_{2},...,d_{K_{\text{s}}-1} \}, \quad d_{k} = t_{k+1} -  t_{k}
        \end{equation}
        If all the intervals are equal ($ \forall k: \,d_{k} = T_{\text{s}}$), then $\mathbb{T}$ is a uniform sampling pattern with a sampling period
        equal to $T_{\text{s}}$.
        If the time intervals are chosen randomly, then $\mathbb{T}$ is a random sampling pattern.

\begin{figure*}[h]
    \includegraphics[scale=1.0]{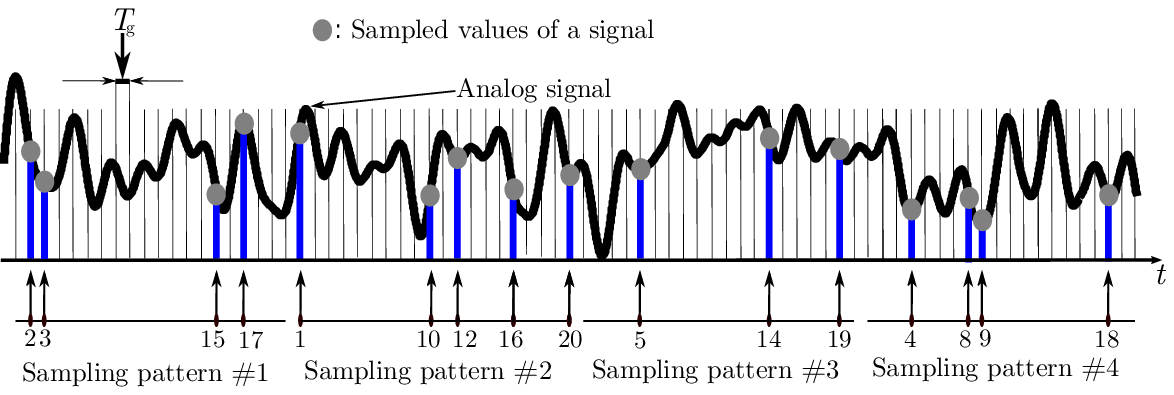}
    \caption{Example of unconstrained random sampling patterns applied to an analog signal.
            There is no minimum nor maximum allowed interval between sampling points.
            Furthermore, patterns contain different number of sampling points.}
    \label{fig:pattern_unconstrained}
\end{figure*}

        A random sampling pattern $\mathbb{T}$ is applied to a signal $s(t)$ of length $\tau$:
        \begin{equation}
         \vect{y}[k] = s(t_{k}), \quad t_{k} \in \mathbb{T}
        \end{equation}
        where $\vect{y} \in \mathbb{R}^{K_{\text{s}}} $ is a vector of observed signal samples.
        The average sampling frequency $f_{\text{s}}$ of a random sampling pattern depends on the number of sampling time points in the pattern:
        \begin{equation}
          f_{\text{s}} = \frac{K_{\text{s}}}{\tau}
        \end{equation}
        An example of a random sampling pattern is shown in Fig. \ref{fig:pattern_unconstrained}.

%2.2
\subsection{Random patterns generation problem}
\label{subsec:patterns_prob}
Let us denote a nontrivial problem $\mathcal{P}(N,\tau,T_{\text{g}},f_{\text{s}}^{\dagger},t_{\text{min}},t_{\text{max}})$
of generation of a multiset (bag) $\mathbb{A}$ with $N$ random sampling patterns.
The time length of sampling patterns is $\tau$, grid period is $T_{\text{g}}$.
The requested average sampling frequency of patterns is $f_{\text{s}}^{\dagger}$,
minimum and maximum intervals between sampling points are $t_{\text{min}}$ and $t_{\text{max}}$ respectively.
The problem $\mathcal{P}$ is solved by random sampling pattern generators.
The generators should meet
requirements given in \ref{subsec:generators_rec},
and all the produced sampling patterns must meet the requirements given below in \ref{subsec:patterns_rec}.

%2.3
\subsection{Requirements for random sampling patterns}
\label{subsec:patterns_rec}

%2.3.1
\subsubsection{Frequency stability}
\label{subsec:patterns_rec_freq}
A random sampling pattern generator must produce sampling patterns
with a requested average sampling frequency $f_{\textbf{s}}^{\dagger}$.
If the average sampling frequency $f_{\textbf{s}}$ is lower than
the requested sampling frequency, then the quality of signal reconstruction
may be compromised.
On the contrary, higher sampling frequency $f_{\textbf{s}}$ than the requested
$f_{\textbf{s}}^{\dagger}$ causes unnecessary power consumption.

%2.3.2
\subsubsection{Minimum and maximum time intervals}
\label{subsec:patterns_rec_dist}
A requirement for minimum interval $t_{\text{min}}$ between sampling points comes from
the ADC technological constraints \cite{An01,Le05,Hui01,Xil01}.
Violation of this requirement may render the sampling pattern impossible to implement with a given ADC.
Similarly, there may be a requirement
of maximum interval between samples $t_{\text{max}}$.
Generating an adequate random sampling pattern is realizable
if $t_{\text{min}} \leq T_{\textbf{s}}^{\dagger}$ and
$t_{\text{max}} \geq T_{\textbf{s}}^{\dagger}$,
where $T_{\textbf{s}}^{\dagger} = 1/f^{\dagger}_{\textbf{s}}$
is the requested average sampling period.

%2.3.3
\subsubsection{Unique sampling points}
As stated in (\ref{eq:mono}), sampling points in
a given sampling pattern $\mathbb{T}$ cannot be repeated.
Repeated sampling points do not make practical sense since a signal
can be sampled only once in a given time moment.
If a sampling pattern contains repeated sampling  points,
then a dedicated routine must remove these repeated points.

%2.4
\subsection{Requirements for random sampling pattern generators}
\label{subsec:generators_rec}

%2.4.1
\subsubsection{Uniform probability density function for grid points}
As described in \ref{subsec:patterns},
a sampling pattern $\mathbb{T}$ is an ordered set
which is a subset of a grid $\mathbb{G}$.
In other words, sampling points are drawn from a pool of grid points.
The sampling pattern generator should not favor any of the sampling grid points.
Ideally, all of the sampling points should be equi-probable.

%2.4.2
\subsubsection{Pattern uniqueness}
Repeated sampling patterns
generate unnecessary processing overhead,
especially if sampling patterns are generated offline
and further processed (Fig. \ref{fig:gen_case2}).
An additional search routine which removes replicas of sampling patterns must be implemented in this case.
Therefore, the ideal random sampling pattern generator should not repeat sampling patterns
unless all the possible sampling patterns have been generated.

%-------------------------------------------------------------------------
%  ----- ----- ----- STATISTICAL EVALUATION
%-------------------------------------------------------------------------
\section{Statistical evaluation of random sampling pattern generators}
\label{sec:parameters}
In this Section we propose statistical parameters for
evaluation of a tested random sampling pattern generator.
Aim of these parameters is to assess how well
sampling patterns produced by the evaluated generator
cope with the requirements described in \ref{subsec:patterns_rec} and \ref{subsec:generators_rec}.
These parameters are to be computed for a bag $\mathbb{A}$ of $N$
patterns produced by the evaluated generator,
the parameters are computed using the Monte Carlo method.
It is checked if every generated sampling pattern
fulfills requirements given in \ref{subsec:patterns_rec}
and if a generated bag (multiset) of sampling patterns
fulfill requirements given in the \ref{subsec:generators_rec}.
According to our best knowledge, similar statistical evaluation has never been
introduced before.

%3.1
\subsection{Frequency stability error parameters}
Let us introduce a statistical parameter
indicating how well the evaluated generator fulfills the imposed requirement
of the requested average sampling frequency $f_{\text{s}}^{\dagger}$ (\ref{subsec:patterns}):
\begin{equation}
  e_{\text{f}} = \\
  \frac{1}{N} \sum_{n=1}^{N}{ \left(  \frac{ f^{\dagger}_{\text{s}}-f^{(n)}_{\text{s}}}{f^{\dagger}_{\text{s}}}  \right )^{2} } = \\
  \frac{1}{N} \sum_{n=1}^{N}{ \left(  \frac{ K_{\text{s}}^{\dagger}-K_{\text{s}}^{(n)}}{K_{\text{s}}^{\dagger}}  \right )^{2} }
\end{equation}
where $f_{\text{s}}^{(n)}$ is the average sampling frequency of the $n$-th sampling pattern.
Since all the sampling patterns have the same time length $\tau$, in practice it is usually
more convenient to use the requested number of sampling points in a pattern $K_{\text{s}}^{\dagger}$
and count the number of actual sampling points in a pattern $K_{\text{s}}^{(n)}$.
This parameter is an average value of a relative frequency error of every sampling pattern.
The lower the parameter $e_{\text{f}}$ is, the better is the frequency stability of the generator.
Additionally, let us introduce a $\gamma_{\text{f}}$ parameter:
\begin{equation}
\gamma_{\text{f}} = \frac{1}{N} \sum_{n=1}^{N}{ \gamma^{(n)}_{\text{f}}  } \quad\quad \gamma^{(n)}_{\text{f}} =
                \begin{cases}
                        0   &\text{for $ K^{\dagger}_{\text{s}} =    K^{(n)}_{\text{s}} $} \\
                        1   &\text{for $ K^{\dagger}_{\text{s}} \neq K^{(n)}_{\text{s}} $} \\
                \end{cases}
\end{equation}
which is the ratio of patterns in a bag $\mathbb{A}$
which violate the frequency stability requirement.
The parameter $\gamma^{(n)}_{\text{f}} = 1$ denotes whether the average sampling frequency of
the $n$-th pattern is incorrect.

%3.2
\subsection{Sampling point interval error parameters}
Let us introduce statistical parameters which
indicate how well the assessed generator meets
the interval requirements discussed in Sec. \ref{subsec:patterns_rec_dist}.
For a given $n$-th sampling pattern $\mathbb{T}^{(n)}$
let us create ordered subsets $\mathbb{D}^{(n)}_{-} \subset \mathbb{D}^{(n)}$
and $\mathbb{D}_{+}^{(n)} \subset \mathbb{D}^{(n)}$,
where $\mathbb{D}$ is a set with intervals between sampling points as in (\ref{eq:distset}).
These subsets contain intervals between samples which violate the minimum
and the maximum requirements between sampling points $t_{\text{min}}$ and $t_{\text{max}}$ respectively:
\begin{equation}
        \mathbb{D}_{-} = \{ d_{-,k}  \in \mathbb{D}:d_{-,k} < t_{\text{min}}  \}
\end{equation}
\begin{equation}
        \mathbb{D}_{+} = \{ d_{+,k}  \in \mathbb{D}:d_{+,k} > t_{\text{max}}  \}
\end{equation}
Now let us introduce statistical parameters $e_{\text{min}}$ and $e_{\text{max}}$:
\begin{equation}
        e_{\text{min}} = \frac{1}{N} \sum_{n=1}^{N}{ (e_{-}^{(n)})^{2} } \quad\quad e_{-}^{(n)} = \frac{ | \mathbb{D}^{(n)}_{-} | }{|\mathbb{D}^{(n)} |}
\end{equation}
\begin{equation}
        e_{\text{max}} = \frac{1}{N} \sum_{n=1}^{N}{ (e_{+}^{(n)})^{2} } \quad\quad e_{+}^{(n)} = \frac{ |\mathbb{D}^{(n)}_{+} | }{|\mathbb{D}^{(n)} |}
\end{equation}
where $|\cdot|$ denotes the number of elements in a set (set's cardinality), and $|\mathbb{D}^{(n)}| = K_{\text{s}}-1$ as in (\ref{eq:distset}).
These parameters contain the average squared ratio
of the number of intervals in a pattern which violate minimum/maximum interval requirements
to the number of all intervals between sampling points in a pattern.
The lower the above parameters are, the better the evaluated generator meets interval requirements.
Similarly to the frequency stability parameter,
let us introduce $\gamma_{\text{min}}$ and $\gamma_{\text{max}}$ parameters:
\begin{equation}
        \gamma_{\text{min}} = \frac{1}{N} \sum_{n=1}^{N}{ \gamma_{\text{min}}^{(n)}  } \quad \quad \\
        \gamma_{\text{min}}^{(n)} =             \begin{cases}
                                              0   &\text{for $ | \mathbb{D}^{(n)}_{-} | = 0 $} \\
                                              1   &\text{for $ | \mathbb{D}^{(n)}_{-} | > 0 $} \\
                                        \end{cases}
\end{equation}
\begin{equation}
        \gamma_{\text{max}} = \frac{1}{N} \sum_{n=1}^{N}{ \gamma_{\text{max}}^{(n)}  } \quad \quad \\
        \gamma_{\text{max}}^{(n)} =             \begin{cases}
                                              0   &\text{for $ | \mathbb{D}^{(n)}_{+} | = 0 $} \\
                                              1   &\text{for $ | \mathbb{D}^{(n)}_{+} | > 0 $} \\
                                        \end{cases}
\end{equation}
which are additional parameters which are equal to ratios of patterns
which violate minimum or maximum intervals between sampling patterns.
Parameters $\gamma_{\text{min}}^{(n)} = 1$ and $\gamma_{\text{max}}^{(n)} = 1$ denote if the $n$-th pattern meets the requirement of minimum and maximum intervals respectively.

%3.3
\subsection{Ratio of incorrect patterns}
It is possible to assign to every $n$-th pattern a parameter $\gamma^{(n)}$
which denotes if a pattern violates the frequency stability (\ref{subsec:patterns_rec_freq}) or
the interval requirements (\ref{subsec:patterns_rec_dist}).
The ratio of incorrect patterns $\gamma$ of a bag $\mathbb{A}$ is:
\begin{equation}
\label{eq:generalError}
    \gamma = \frac{1}{N} \sum_{n=1}^{N}{ \gamma^{(n)}  } \quad\quad \gamma^{(n)} = \gamma^{(n)}_{\text{f}} \;\vee\; \gamma^{(n)}_{\text{min}} \;\vee\; \gamma^{(n)}_{\text{max}}
\end{equation}
where $\vee$ is a logical disjunction.
Using parameter $\gamma^{(n)}$  it is possible to generate a sub-bag $\mathbb{A}^{\star} \sqsubseteq \mathbb{A}$
which contains only correct patterns from the bag $\mathbb{A}$:
\begin{equation}
\label{eq:bagAstar}
     \mathbb{A}^{\star} = \{ \mathbb{T} \;\; \textbf{in} \;\; \mathbb{A} : \;\; \gamma^{(n)} = 0 \}
\end{equation}
where $\mathbb{T} \;\; \textbf{in} \;\; \mathbb{A}$ signifies that a pattern $\mathbb{T}$ is an element of a multiset $\mathbb{A}$.
Please note that $\mathbb{A}$ is a multiset, so patterns which are the elements of $\mathbb{A}$ may be repeated,
and patterns which are the elements of the multiset $\mathbb{A}^{\star}$ may also be repeated.
Ideally, a sub-bag with correct patterns $\mathbb{A}^{\star}$ is identical to the original bag $\mathbb{A}$.

%3.4
\subsection{Quality parameter: Probability density function}
\label{sec:qPDF}
Let us introduce a statistical parameter $e_{\text{p}}$ which indicates whether
the probability density of occurrence for grid points in patterns from bag $\mathbb{A}$ is uniformly distributed:
\begin{equation}
\label{eq:PDFpar}
e_{\text{p}} = \frac{1}{K_{\text{g}}}\sum_{m=1}^{K_{\text{g}}}{(p_{\text{g}}(m)-1)^2}
\end{equation}
The probability of occurrence of the $m$-th grid point $p_{\text{g}}(m)$ is:
\begin{equation}
p_{\text{g}}(m) = \frac{K_{\text{g}}}{K_{\text{t}}}\sum_{n=1}^{N}{\text{g}_{m}(n)} \quad\quad K_{\text{t}} = \sum^{N}_{n}K_{\text{s}}^{(n)}
\end{equation}
where $K_{\text{g}}$ is the number of sampling grid points in a sampling pattern,
$K_{\text{t}}$ is the total number of sampling points in all the patterns in a bag $\mathbb{A}$,
and the parameter $\text{g}_{m}(n)$ indicates whether the $m$-th grid point is used in the $n$-th sampling pattern $\mathbb{T}^{(n)}$:
\begin{equation}
\text{g}_{m}(n) = \\
\begin{cases}
 0 & \text{if  } mT_{\text{g}} \notin \mathbb{T}^{(n)} \\
 1 & \text{if  } mT_{\text{g}} \in \mathbb{T}^{(n)}
\end{cases}
\end{equation}
Additionally,
let us introduce a statistical parameter $e_{\text{p}}^{\star}$ which
is calculated identically to $e_{\text{p}}$, but based on
sampling patterns from subbag $\mathbb{A}^{\star}$ (\ref{eq:bagAstar}).

%3.5
\subsection{Quality parameter: Uniqueness of patterns}
\label{sec:qU}
Let us create a set $\mathbb{A}_{\#}$
for a bag $\mathbb{A}$ of $N$ sampling patterns generated by the evaluated pattern generator
which contains only unique patterns from $\mathbb{A}$.
Similarly, let us create a set $\mathbb{A}_{\#}^{\star}$
which contains only unique patterns from the subbag with correct patterns $\mathbb{A}^{\star}$ (\ref{eq:bagAstar}).
Now let us introduce parameters $\eta_N$ and $\eta^{\star}_N$:
\begin{equation}
\label{eq:uniq}
\eta_N = |\mathbb{A}_{\#}| \quad\quad \eta^{\star}_N = |\mathbb{A}^{\star}_{\#}|
\end{equation}
These parameters count the number of unique patterns and unique correct patterns in the bag $\mathbb{A}$ with $N$ generated patterns.

%-------------------------------------------------------------------------
%  ----- ----- ----- PATTERNS GENERATORS
%-------------------------------------------------------------------------
\section{Pattern generators}
\label{sec:generators}
Algorithms of sampling pattern generators are presented in this Section.
Subsection \ref{sec:JS_ARS} presents existed, widely known Jittered Sampling (JS) and Additive Random Sampling (ARS) algorithms.
Subsection \ref{sec:ANGIE} presents the proposed sampling pattern generator algorithm, which is tailored to
fulfill the requirements presented in \ref{subsec:patterns_rec} and \ref{subsec:generators_rec}.
Please note that all the algorithms presented in this paper generate sampling patterns
represented as indices of sampling grid points as in (\ref{eq:grid_representation}).

%4.1
\subsection{Jittered Sampling and Additive Random Sampling}
\label{sec:JS_ARS}
Jittered Sampling and Additive Random Sampling algorithms are widely used to generate random sequences.
There are 4 input variables to the JS and ARS algorithms:
requested time of a sampling pattern $\tau$,
grid period $T_{\text{g}}$,
requested average sampling frequency $f^{\dagger}_{\text{s}}$ and
the variance parameter $\sigma^{2}$.
The realizable time of a sampling pattern $\hat{\tau}$ may differ from the given requested time of a pattern $\tau$ if the given time is not a multiple of the given grid period $T_{\text{g}}$.
Before either of the algorithms is started,
the number of grid points $K_{\text{g}}$ in a sampling pattern,
the realizable time of a sampling pattern $\hat{\tau}$
and the realizable requested number of sampling points $\hat{K}^{\dagger}_{\text{s}}$ must be computed:
\begin{equation}
\label{eqalg:precomp1}
K_{\text{g}} = \left \lfloor \frac{\tau}{T_{\text{g}}}  \right \rfloor \quad\quad
\hat{\tau} = K_{\text{g}} T_{\text{g}} \quad\quad
\hat{K}^{\dagger}_{\text{s}} = [ \hat{\tau} f^{\dagger}_{\text{s}}]
\end{equation}
where $[\cdot]$ signifies the rounding function, which returns an integer which is closest to the function's argument.
Because the algorithms operate on a discrete set of grid points,
the realizable requested average sampling frequency $\hat{f}^{\dagger}_{\text{s}}$
may differ from the requested sampling frequency $f^{\dagger}_{\text{s}}$.
The realizable requested average sampling frequency $\hat{f}^{\dagger}_{\text{s}}$
and realizable requested average sampling period $\hat{T}^{\dagger}_{\text{s}}$ is computed:
\begin{equation}
\label{eqalg:hatsf}
\hat{f}^{\dagger}_{\text{s}} = \frac{\hat{K}^{\dagger}_{\text{s}}}{\hat{\tau}} \quad\quad
\hat{T}^{\dagger}_{\text{s}} = \frac{1}{\hat{f}^{\dagger}_{\text{s}}} \quad\quad
\hat{N}^{\dagger}_{\text{s}} = \left [  \frac{\hat{T}^{\dagger}_{\text{s}}}{T_{\text{g}}} \right ]
\end{equation}
where $\hat{N}^{\dagger}_{\text{s}}$ is the requested average sampling period recalculated to the number of grid periods.
If the computed realizable requested sampling frequency $\hat{f}^{\dagger}_{\text{s}}$ is different from the requested sampling frequency $f^{\dagger}_{\text{s}}$,
the problem of generation of sampling patterns is not well stated.
Before the algorithms start, the index of a correct sampling point $\hat{k}$ and the starting position of the sampling point $n_0$ must be reset:
\begin{equation}
\label{eqalg:ARSJSreset}
\hat{k} = 0 \quad n_{0} = 0
\end{equation}

\begin{figure*}[h]
   \centering
    \includegraphics[scale=1.0]{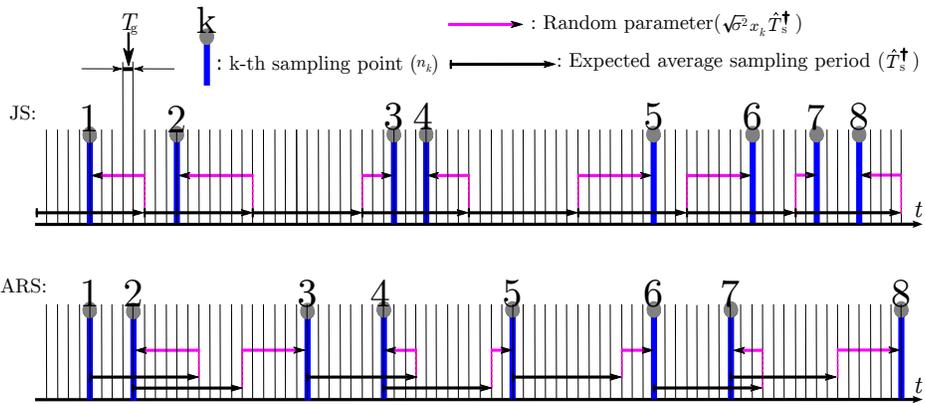}
    \caption{Illustration of generation of sampling patterns in Jittered Sampling (JS) and Additive Random Sampling (ARS) algorithms.}
    \label{fig:ARS_JS_illustration}
\end{figure*}

In the JS algorithm, every sampling point is a uniform sampling point which is randomly "jittered":
\begin{equation}
\label{eqalg:drawJS}
n_{\mathrm{JS},k}^{\ast} = [k \hat{N}^{\dagger}_{\text{s}} + \sqrt{\sigma^{2}} x_k \hat{N}^{\dagger}_{\text{s}}]\quad x_k \thicksim \mathcal{N}(0,1)
\end{equation}
where $\mathcal{N}(0,1)$ denotes a standard normal distribution.   
In the ARS algorithm every sampling point is computed using the previous sampling point
to which an average sampling period and a random value are added:
\begin{equation}
\label{eqalg:drawARS}
n_{\mathrm{ARS},k}^{\ast} = [n_{\hat{k}-1} + \hat{N}^{\dagger}_{\text{s}} + \sqrt{\sigma^{2}} x_k \hat{N}^{\dagger}_{\text{s}}] \quad x_k \thicksim \mathcal{N}(0,1)
\end{equation}
Fig. \ref{fig:ARS_JS_illustration} illustrates generation of sampling patterns in the JS and ARS algoritms.

The practical versions of both JS and ARS algorithms are presented in Alg. \ref{alg:JSARS}.
After generation of a pattern, any repeated sampling point must be removed (line 12 of Alg. \ref{alg:JSARS}).
It is because in these algorithms there is no guarantee that sampling points are not repeated.
\begin{algorithm}[htbp]
        \caption{JS and ARS algorithms - pseudo code}
        \label{alg:JSARS}
        \begin{algorithmic}[1]
                \STATE {\color{red}{\bf function} $[ \bm{\mathbb{T}} ]  =  \mbox{\tt{JS/ARS}}(\bm{\tau}, \bm{T_{\text g}}, \bm{f_{\text s}^{\dagger}}, \bm{\sigma^{2}})$}
                \STATE Compute $K_{\text{g}}$, $\hat{\tau}$ and $\hat{K}^{\dagger}_{\text{s}}$ as in (\ref{eqalg:precomp1})
                \STATE Compute $\hat{f}^{\dagger}_{\text{s}}$, $\hat{T}^{\dagger}_{\text{s}}$ and $\hat{N}^{\dagger}_{\text{s}}$ as in (\ref{eqalg:hatsf})
                \STATE Reset $\hat{k}$ and $n_{0}$ as in (\ref{eqalg:ARSJSreset})

                \STATE FOR $k = 1$ TO $\hat{K}^{\dagger}_{\text{s}}$
                \STATE \idt Draw sampling moment $n_{\mathrm{JS}, k}^{\ast}$ (\ref{eqalg:drawJS}) or $n_{\mathrm{ARS}, k}^{\ast}$ (\ref{eqalg:drawARS})
                \STATE \idt IF $n_k^{\ast} > 0$ AND $n_k^{\ast} < \hat{\tau}$
                \STATE \idt \idt $n_{\hat{k}} \leftarrow n_{\hat{k}}^{\ast}$
                \STATE \idt \idt Assign $\mathbb{T}'(\hat{k}) \leftarrow n_{\hat{k}}$
                \STATE \idt \idt $\hat{k} \leftarrow \hat{k} + 1$
                \STATE END
                \STATE Remove repeated sampling points in $\mathbb{T}$
        \end{algorithmic}
\end{algorithm}

%4.2
\subsection{'ANGIE' algorithm}
\label{sec:ANGIE}
We propose an algorithm which would
perfectly cope with the requirements described in \ref{subsec:patterns_rec}
and as much as possible with the requirements in \ref{subsec:generators_rec}.
The ratio of incorrect patterns $\gamma$ (\ref{eq:generalError}) generated by the algorithm should always equal 0,
while keeping the probability density parameter $e_{\text{p}}$ (Sec. \ref{sec:qPDF}) as low as possible and the uniqueness parameter $\eta_N = \eta_N^{\star}$ (Sec. \ref{sec:qU}) as high as possible.
The parameters $e^{\star}_{\text{p}}$ and $\eta_N^{\star}$ must equal $e_{\text{p}}$ and $\eta_N$ respectively,
as the subbag with correct patterns $\mathbb{A}^{\star}$ must be identical to the subbag with all the patterns $\mathbb{A}$ (all the generated patterns must be correct).
Therefore we propose the rANdom sampling Generator with Intervals Enabled (ANGIE) algorithm.
The input variables to the algorithm are identical to the JS and ARS algorithms (\ref{sec:JS_ARS}),
with additional variables for the allowed time between samples ($t_{\text{min}}$, $t_{\text{max}}$).

Before the ANGIE algorithm starts, the following precomputations must be done.
Similarly to the JS and ARS algorithms,
the number of grid points in a sampling pattern ($K_{\text{g}}$),
the realizable time of a sampling pattern ($\hat{\tau}$) and
the realizable number of sampling points in a sampling pattern $\hat{K}^{\dagger}_{\text{s}}$ must be computed as in (\ref{eqalg:precomp1}).
Then the minimum and the maximum time between sampling points must be recalculated to the number of grid points:
\begin{equation}
\label{eqalg:2ndPrep}
K_{\text{min}} = \left \lceil \frac{t_{\text{min}}}{T_{\text{g}}} \right \rceil \quad K_{\text{max}} = \left \lfloor \frac{t_{\text{max}}}{T_{\text{g}}} \right \rfloor
\end{equation}
where $\left \lceil \cdot \right \rceil$ signifies the ceiling function which returns the lowest integer which is higher or equal to the function's argument.
In the proposed algorithm there are 2 limit variables, $n_k^-$ and $n_k^+$, which are the first and the last possible position of a $k$-th
sampling point. These variables are updated after generation of every sampling point.
Before the algorithm starts these variables must be initialized:
\begin{equation}
\label{eqalg:ResetLimits}
n_1^- = 1 \quad\quad n_1^+ = K_{\text{g}} - K_{\text{min}}(\hat{K}^{\dagger}_{\text{s}}-1)
\end{equation}
The number of sampling points left to be generated is updated before generation of every sampling point:
\begin{equation}
\label{eqalg:samp_left}
n_{k}^{\text{left}} = \hat{K}^{\dagger}_{\text{s}} - k + 1
\end{equation}
where $k$ is the index of the current sampling point.
The average sampling period for the remaining $n_{k}^{\text{left}}$ sampling points and
the expected position $e_{k}$ of the $k$-th sampling point is:
\begin{equation}
\label{eqalg:avg_samp_period}
e_{k} = n_{k-1} + n_k^{\ddagger} \quad\quad n_k^{\ddagger} = \left [ \frac{K_{\text{g}} - n_{k-1}}{n_{k}^{\text{left}} + 1} \right ]
\end{equation}
In the proposed algorithm, a $k$-th sampling point $n_k$ may differ from its expected position $e_{k}$ by the interval
$n_k^{\text{d}}$.
Before computing this interval the algorithm must compute intervals to the limits:
\begin{equation}
n_k^{\text{d}-} = | e_{k} - n_k^{-}| \quad\quad n_k^{\text{d}+} = | n_k^{+} - e_{k}|
\end{equation}
and then the lower from the above intervals is the correct interval $n_k^{\text{d}}$:
\begin{equation}
\label{eqalg:dist}
n_k^{\text{d}} = \min{(n_k^{\text{d}-},n_k^{\text{d}+})}
\end{equation}

The first sampling point is drawn using a uniformly distributed variable $x^{u}$:
\begin{equation}
\label{eqalg:sampmom_uniq}
  n_1 = \lceil x_1^u n_k^{\ddagger} \rceil \quad x_1^u \thicksim \mathcal{U}(0,1)
\end{equation}
while the rest of the sampling points are drawn using the normal distribution:
\begin{equation}
\label{eqalg:sampmom_norm}
  n_k = e_k + [ x_k n_k^{\text{d}}]\quad x_k \thicksim \mathcal{N}(0,\sigma^{2})
\end{equation}
Finally, the algorithm checks whether the drawn sampling moment $n_k$
violates the limits $n_k^-$ and $n_k^+$:
\begin{equation}
\label{eqalg:check_delim}
        n_k =
                \begin{cases}
                        n^{+}_{k}   &\text{for $ n_k > n^{+}_{k} $} \\
                        n^{-}_{k}   &\text{for $ n_k < n^{-}_{k} $} \\
                \end{cases}
\end{equation}
In the last step the limits for the next sampling point are computed.
The lower and the higher limits are computed as:
\begin{equation}
\label{eqalg:delimit_min}
  n_{k+1}^- = n_k + K_{\text{min}} \quad\quad n_{k+1}^+ = K_{\text{g}} - K_{\text{min}}(n_{k}^{\text{left}}-2)
\end{equation}
If the maximum time between samples is valid ($t_{\text{max}} < \inf$),
then the higher limit should be additionally checked for $t_{\text{max}}$:
\begin{equation}
\label{eqalg:delimit_max}
  n_{k+1}^+ = \min{({n_{k+1}^+,n_k + K_{\text{max}}})}
\end{equation}

The proposed algorithm is presented in Alg. \ref{alg:ANGIE}.
\begin{algorithm}[htbp]
        \caption{'ANGIE' algorithm - pseudo code}
        \label{alg:ANGIE}
        \begin{algorithmic}[1]
                \STATE {\color{red}{\bf function} $[ \bm{\mathbb{T}} ]  =  \mbox{\tt{ANGIE}}(\bm{\tau}, \bm{T_{\text g}}, \bm{f_{\text s}^{\dagger}}, \bm{t_{\text{min}}}, \bm{t_{\text{max}}}, \bm{\sigma^{2}})$}

                \STATE Compute $K_{\text{g}}$, $\hat{\tau}$ and $\hat{K}^{\dagger}_{\text{s}}$ as in (\ref{eqalg:precomp1})

                \STATE Compute $K_{\text{min}}$ and $K_{\text{max}}$ as in (\ref{eqalg:2ndPrep})
                \STATE Initialized the limits $n_1^-$ and $n_1^+$ as in (\ref{eqalg:ResetLimits})
                \STATE FOR $k = 1$ TO $\hat{K}^{\dagger}_{\text{s}}$
                \STATE \idt Update the number of sampling points left $n^{\text{left}}_{k}$ as in (\ref{eqalg:samp_left})

                \STATE \idt Compute the expected position $e_{k}$ as in (\ref{eqalg:avg_samp_period})

                \STATE \idt Compute the interval $n_{k}^{\text{d}}$ as in (\ref{eqalg:dist})

                \STATE \idt Draw sampling moment $n_{k}$ as in (\ref{eqalg:sampmom_uniq}) or (\ref{eqalg:sampmom_norm})

                \STATE \idt Check and correct $n_{k}$ as in (\ref{eqalg:check_delim})

                \STATE \idt Assign $\mathbb{T}'(k) \leftarrow n_{k}$

                \STATE \idt Update the limits $n_{k+1}^{-}$ and $n_{k+1}^{+}$ as in (\ref{eqalg:delimit_min}) and (\ref{eqalg:delimit_max})
                \STATE END
        \end{algorithmic}
\end{algorithm}

%-------------------------------------------------------------------------
%  ----- ----- ----- NUMERICAL EXPERIMENTS
%-------------------------------------------------------------------------
\section{Numerical experiment}
\label{sec:taa}
In this section, the performance of the proposed ANGIE algorithm
is experimentally compared with the JS and ARS algorithms.
A toolbox with pattern generators and evaluation functions was created to
facilitate the experiment. Emphasis was set on validation of parts of the software.
The toolbox, together with its documentation, is available online at \cite{Jap01}.
Using the content available at \cite{Jap01} it is possible to reproduce the presented
numerical simulations.

%5.1
\subsection{Experiment \#1 - setup}
\label{sec:expsetup}
The duration $\tau$ of sampling patterns is set to 1 ms, sampling grid period $T_{\text{g}}$
is equal to 1 $\mu$s.
The requested average sampling frequency of patterns is set to 100 $\kHz$,
which corresponds to an average sampling period equal to 10 $\mu$s.
The minimum time between sampling points is $t_{\text{min}} = 5 \mu$s,
and there is no requirement for maximum time between sampling points ($t_{\text{max}} = \inf$).
The variance $\sigma^{2}$ is logarithmically swept in the range $[10^{-4}, 10^2]$.

The computed statistical parameters of sampling patterns are automatically tested for convergence.
A mean value is accounted as converged, if for the last $2 \cdot 10^{4}$ patterns
it did not change more than 1\% of the mean value computed for all the patterns currently tested.
The minimum number of sampling patterns tested is $10^{5}$.
The uniqueness parameters $\eta_{N}$ and $\eta_{N}^{\star}$ (\ref{eq:uniq}) are computed after $N = 10^{5}$ patterns.

%5.2
\subsection{Experiment \#1 - results }
\begin{figure*}[h]
    \centering
    \includegraphics[scale=1.0]{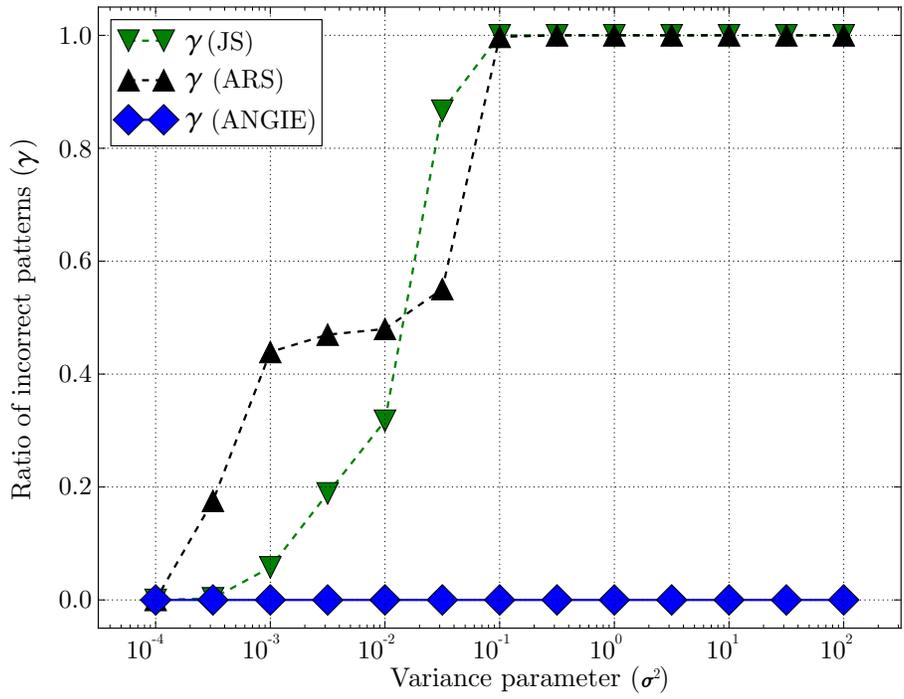}
    \caption{Ratio of incorrect patterns $\gamma$ computed for patterns generated by the JS, ARS and ANGIE algorithms (experiment \#1).}
    \label{fig:errratio}
\end{figure*}

\begin{figure*}[h]
    \centering
    \includegraphics[scale=1.0]{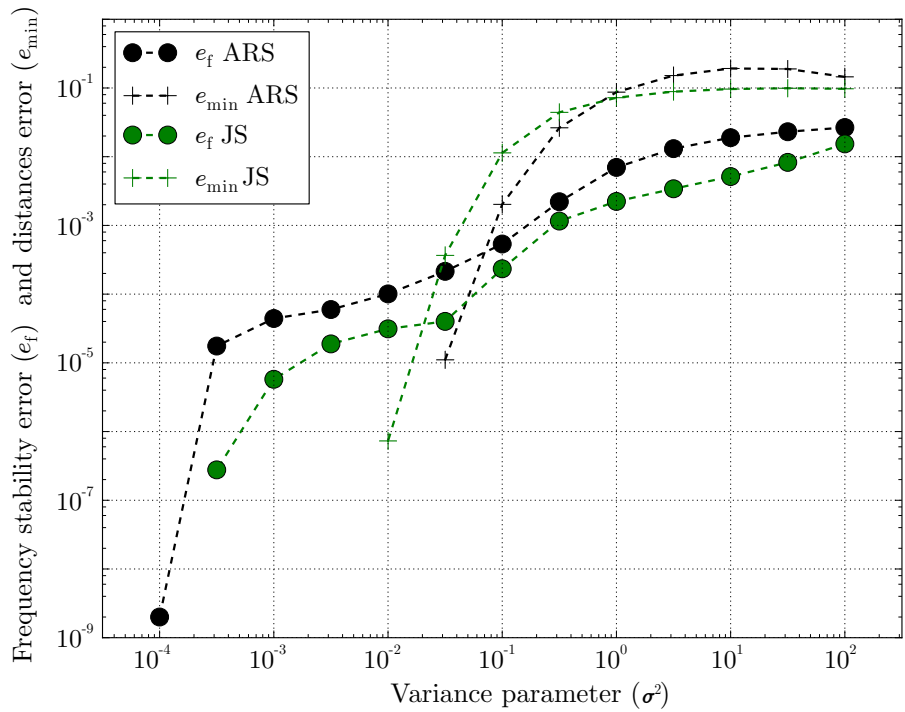}
    \caption{Frequency stability error $e_{\text{f}}$ and intervals error $e_{\text{min}}$ computed for patterns generated by the JS and ARS algorithms (experiment \#1).
    The error parameters are not plotted for
    the ANGIE algorithm because errors for this algorithm are equal 0 (all the patterns generated by the algorithm are correct - Fig. \ref{fig:errratio}).}
    \label{fig:errparam}
\end{figure*}

\begin{figure*}[h]
    \centering
    \includegraphics[scale=1.0]{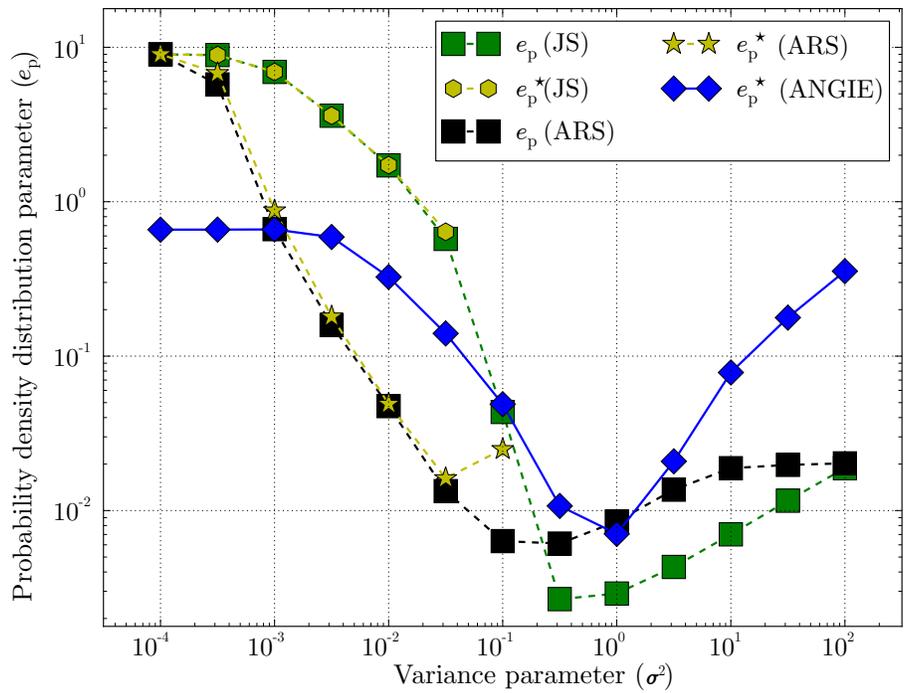}
    \caption{Probability density parameter $e_{\text{p}}$ computed for patterns generated
             by the JS, ARS and ANGIE algorithms (experiment \#1).}
    \label{fig:qualparam}
\end{figure*}

\begin{figure*}[h]
    \centering
    \includegraphics[scale=1.0]{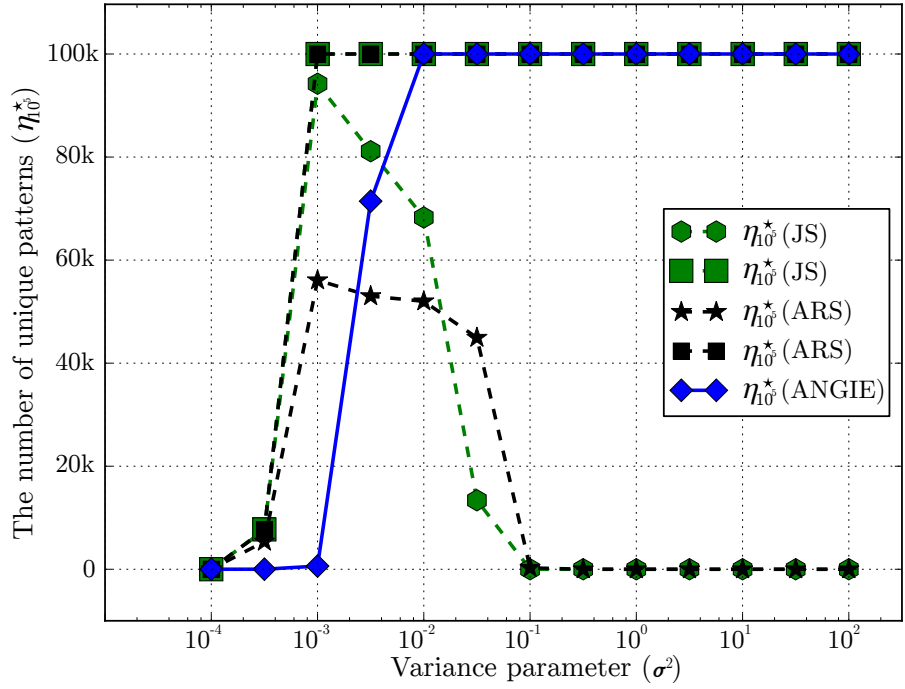}
    \caption{The number of unique patterns $\eta_{10^5}$ computed for patterns generated
             by the JS, ARS and ANGIE algorithms (experiment \#1).
             The parameter $\eta_{10^5}^{\star}$ is not plotted for the ANGIE algorithm since it is equal
             to the parameter $\eta_{10^5}$ for this algorithm.
    It is because the subbag $\mathbb{A}^{\star} = \mathbb{A}$ for the ANGIE algorithm (all the patterns generated by the algorithm are correct - ref. to Fig. \ref{fig:errratio})}
   \label{fig:uniqueparam}
\end{figure*}

Error parameters computed for the tested sampling pattern generators are plotted in Fig. \ref{fig:errparam}.
The ratio of incorrect patterns are plotted in Fig. \ref{fig:errratio}.
This ratio for the ANGIE algorithm (blue $\diamond$) is equal to 0 for all the values of variance $\sigma^{2}$.
Thus, all the pattens have correct average sampling frequency and intervals between sampling points.
Patterns generated by the JS (green $\blacktriangledown$) and the ARS algorithms (black $\blacktriangle$) are all correct for very low values of the variance $\sigma^{2}$,
but the quality parameters $e_{\text{p}}$ and $\eta_{10^{5}}$ for these $\sigma^{2}$ values are poor (Fig. \ref{fig:qualparam} and Fig. \ref{fig:uniqueparam}).
In Fig. \ref{fig:errparam} it can be seen that for nearly all the values of variance $\sigma$,
the frequency stability of the patterns generated by the JS and the ARS algorithms is compromised,
and for most of the values of $\sigma^{2}$, the requirement of minimum intervals between sampling points
is not met by these algorithms.

\begin{figure*}[h]
    \centering
    \includegraphics[scale=1.0]{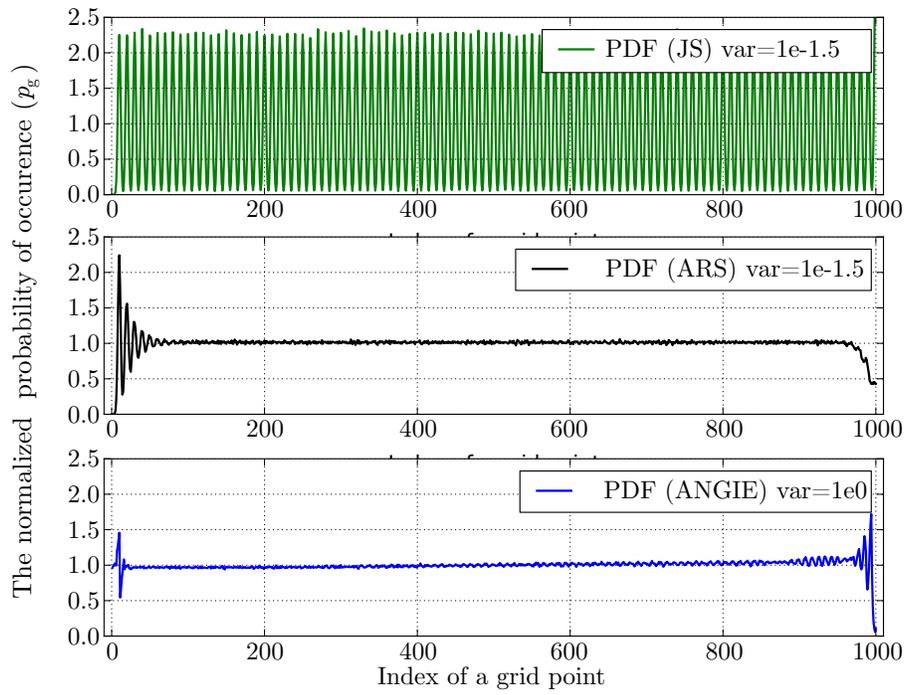}
    \caption{The best probability density functions of grid points found for the tested sampling pattern generators (experiment \#1).}
    \label{fig:plotPDF}
\end{figure*}

The best values of the parameter $e_{\text{p}}$ are achieved for the JS (green $\blacksquare$) and the ARS (black $\blacksquare$) algorithms (Fig. \ref{fig:qualparam}),
but only if all the patterns (also incorrect) are taken into account (parameter $e_{\text{p}}$).
If the quality parameter was computed only for the correct patterns (parameter $e_{\text{p}}^{\star}$),
it can be clearly seen that the proposed algorithm (blue $\blacksquare$) performs significantly
better than the JS (yellow $\hexagon$) and the ARS algorithms (yellow $\star$).
Furthermore, the best values of $e_{\text{p}}^{\star}$ are found for the values of variance $\sigma$
for which most of the patterns produced by the JS and the ARS algorithms are incorrect.
Plots of the best probability density functions found for the tested algorithms are in Fig. \ref{fig:plotPDF}.

Fig. \ref{fig:uniqueparam} shows the number of unique patterns produced by the tested algorithms.
The number of unique correct patterns produced by the proposed algorithm is
higher than the number produced by the JS and the ARS algorithms for any variance value $\sigma^{2} \geq 10^{-2}$.

The above results show that the proposed algorithm ANGIE performs better than the JS and the ARS algorithms.
All the patterns generated by the ANGIE algorithm are correct,
have a parameter $\gamma^{(n)}$ defined as in (\ref{eq:generalError}) equal to 0.
The quality parameters described in Sec. \ref{sec:qPDF} and Sec. \ref{sec:qU} are better for the proposed algorithm.
It can be seen that the variance value $\sigma^{2}$, which is an internal algorithm parameter, should be adjusted to a given problem.
For the given problem, the proposed algorithm performs best for $\sigma^{2} = 10^{-2}$.

\subsection{Experiment \#2 - setup}
In the second experiment four different cases (A-D) of sampling patterns are studied.
Parameters of these cases are collected in Table 1.
In the first two cases there are requirements of both the minimum and the maximum distance between sampling points.
In the second case there are only 5 sampling points requested p. sampling pattern, and the number of sampling grid points is limited to 100.
In the third case there are no requirements imposed on distances between sampling points, so there is only the requirement of stable average sampling frequency.
This case is distinctive from others, because the number of sampling points p. sampling pattern is high ($10^{4}$), and the grid period is very low.
In the last case there is a requirement of the maximum distance between sampling points.
In all the four cases the variance $\sigma^{2}$ is logarithmically swept in the range $[10^{-4}, 10^2]$.

In this experiment there are three quality parameters measured for all the three generators (JS, ARS and ANGIE).
The first parameter is the ratio of incorrect patterns $\gamma$ (\ref{eq:generalError}).
The second is the probability density parameter $e_{\text{p}}^{\star}$ as in (\ref{eq:PDFpar}), but computed only for the correct patterns.
The third quality parameter is the number of unique correct patterns in the first $10^4$ generated patterns $\eta^{\star}_{10^{4}}$ (\ref{eq:uniq}).

        \begin{table}[h!]
        \label{table:exp2}
        \centering
          \begin{tabular}{ c || c | c | c | c | c ||  c | c | c | c}
                 & \multicolumn{5}{ c|| }{Independent parameters}                                       & \multicolumn{4}{ c }{Dependent parameters} \\
                 & $\tau$      & $T_g$           & $f_s$        & $t_{\text{min}}$ & $t_{\text{max}}$  &    $K_g$  & $\hat{K}^{\dagger}_{\text{s}}$ & $K_{\text{min}}$ & $K_{\text{max}}$ \\
            case & \text{[ms]} & \text{[$\mu$s]} & \text{[kHz]} & \text{[ms]}      & \text{[ms]}       &           &                                &                  &  \\
            \hline \\ [-2.2ex]
            A  & $10^3$  & $10^3$                & 0.05         & 10     & 30                & $10^3$           &  50    & 10    & 30 \\
            B  & 0.1     & 1                     & 50           & 0.015  & 0.028             & 100              &   5    & 15    & 28 \\
            C  & $10^3$  & 1                     & 10           & ---    & ---               & $10^6$           & $10^4$ & ---  & --- \\
            D  & 0.005   & 25$\cdot$ $10^{-5}$   & $10^5$       & ---    & 14$\cdot 10^{-6}$ & 2$\cdot$ $10^4$  &  500   & ---  & 56  \\
            \hline
          \end{tabular}
          \caption{Parameters of sampling patterns used in all the four cases of experiment \#2.
          Independend parameters are: time length of sampling patterns ($\tau$), grid period ($T_g$), requested average sampling frequency ($f_s$), minimum allowed time between sampling points ($t_{\text{min}}$),
          maximum allowed time between sampling points ($t_{\text{max}}$). Shown dependent parameters are: the number of grid points ($K_g$), the requested realizable number of sampling points ($\hat{K}^{\dagger}_{\text{s}}$),
          the minimum and maximum time between the sampling points recalculated to the number of grid points ($K_{\text{min}}$ and $K_{\text{max}}$).}
        \end{table}

\subsection{Experiment \#2 - results }
\begin{figure*}[h]
    \centering
   \includegraphics[scale=1.0]{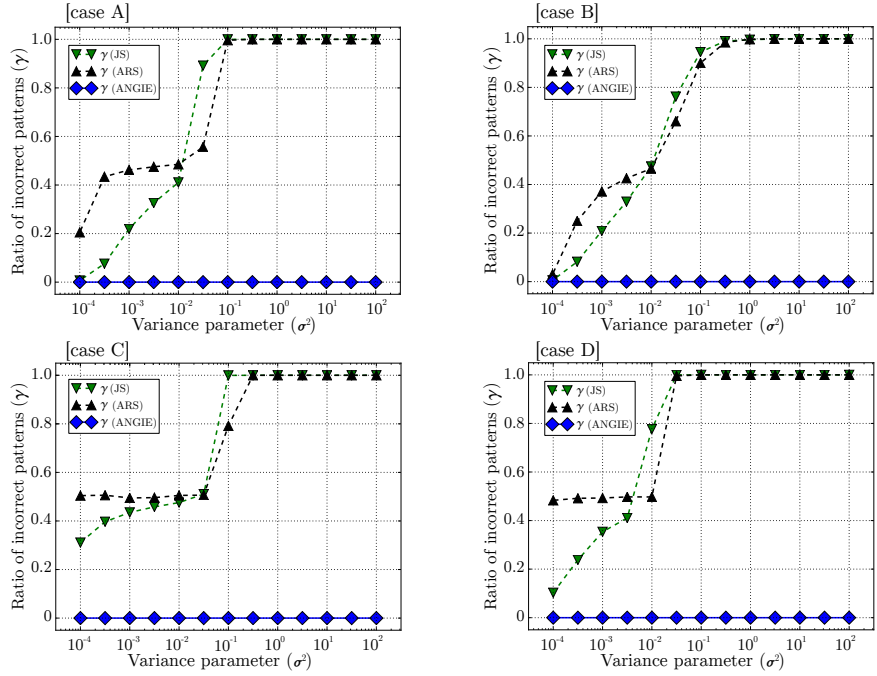}
    \caption{Ratio of incorrect patterns $\gamma$
             computed for patterns generated by the JS, ARS and ANGIE algorithms
             in all the four cases of the experiment \#2.}
    \label{fig:eps2corr}
\end{figure*}

\begin{figure*}[h]
    \centering
    \includegraphics[scale=1.0]{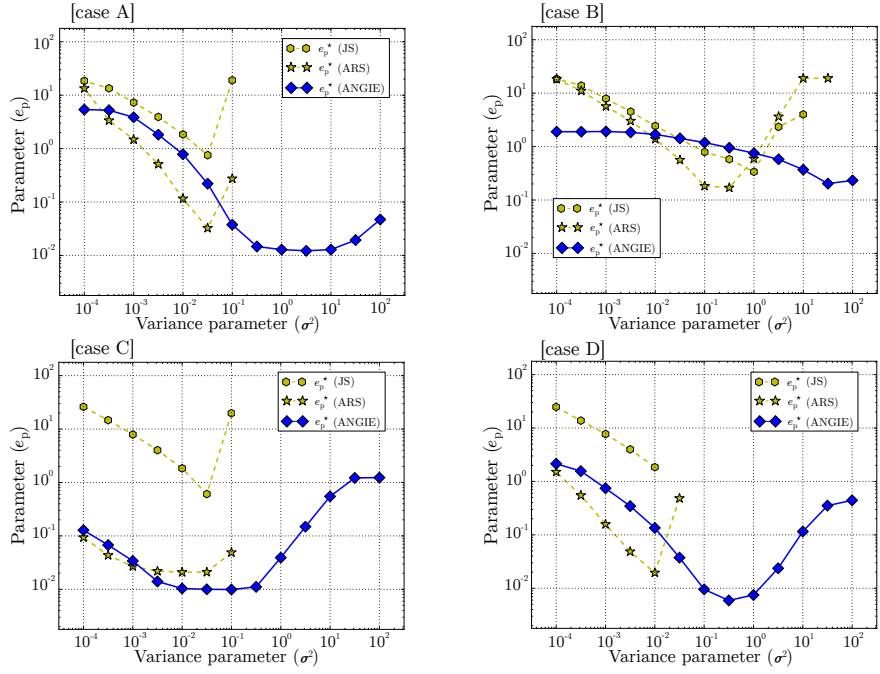}
    \caption{Probability density parameter $e^{\star}_{\text{p}}$ (parameter computed for correct patterns only)
             computed for patterns generated by the JS, ARS and ANGIE algorithms
             in all the four cases of the experiment \#2.}
    \label{fig:exp2ep}
\end{figure*}

\begin{figure*}[h]
    \centering
    \includegraphics[scale=1.0]{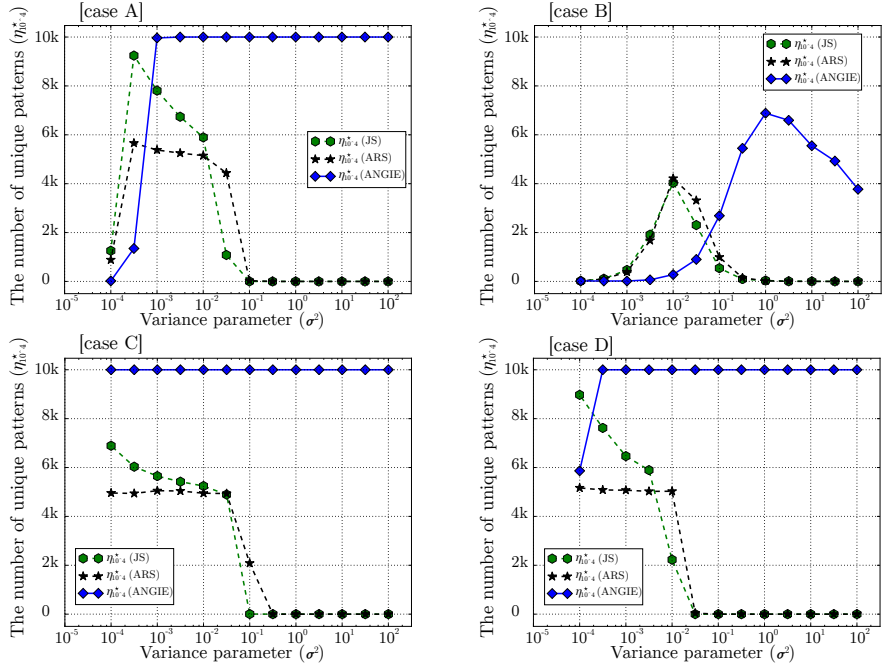}
    \caption{The number of unique patterns $\eta^{\star}_{10^4}$ (parameter computed for correct patterns only)
             computed for patterns generated by the JS, ARS and ANGIE algorithms
             in all the four cases of the experiment \#2.}
    \label{fig:exp2unique}
\end{figure*}

Results of this experiment are shown on Figures \ref{fig:eps2corr}--\ref{fig:exp2unique}. Each Figure presents a measured quality parameter for all the four cases.
The ratio of incorrect patterns $\gamma$ is on Fig. \ref{fig:eps2corr}, the probability density parameter $e_{\text{p}}^{\star}$ is on Fig. \ref{fig:exp2ep},
and the number of unique correct patterns $\eta^{\star}_{10^{4}}$ is on Fig. \ref{fig:exp2unique}.

Let us take a look at the ratio of incorrect patterns (Fig. \ref{fig:eps2corr}).
The ANGIE algorithm generates only correct sampling patterns.
Hence to line 10 in the algorithm (see Algorithm 2),
the minimum and the maximum distances between sampling points are kept.
Lines 6--8 in the ANGIE algorithm ensure that there will be place for the correct number of sampling points in all the generated sampling patterns.
To the contrary, both ARS and JS algorithms generate a lot of incorrect patterns.
For the high values of variance $\sigma^{2}$ there are only incorrect patterns generated by these two algorithms.

In the three cases (A, C, D) the best probability density parameter $e_{\text{p}}^{\star}$ (Fig. \ref{fig:exp2ep})
measured for patterns generated by the ANGIE algorithm is better than for the other two algorithms.
Additionally, it can be seen in Fig. \ref{fig:exp2unique} that the generated number of unique correct sampling patterns is in
all the four cases significantly higher for the proposed ANGIE algorithm.
Let us take a closer look on the case B.
In this case, the best probability density parameter $e_{\text{p}}^{\star}$ found for the algorithm ARS ($\sigma^2 = 10^{-0.5}$)
is slightly better than the best $e_{\text{p}}^{\star}$ found for the ANGIE ($\sigma^2 = 10^{1.5}$).
Still, the number of unique patterns is significantly better for the above values of $\sigma^2$
for ANGIE algorithm, and very most of the patterns generated by the ARS are incorrect for $\sigma^2 = 10^{-0.5}$.

We tried to find a case for which ARS and JS algorithms would clearly and distinctly outperform the ANGIE, but it turned out to be an impossible task.
Still though, it is difficult to provide the reader with one gold rule which algorithm should be used.
In practical applications there may be a huge number of different sampling scenarios,
in this paper we covered only a tiny fraction of examples, and therefore every case should be considered separately.
In general, ANGIE algorithm will always generate correct sampling patterns.
But if these sampling patterns will have all quality parameters (especially $e_{\text{p}}^{\star}$) better than sampling patterns generated by the other
algorithms, that is an another issue.
From our experience we claim that indeed, in most of the cases ANGIE is the right choice.
However, there might be applications in which, for example, equi-probability of occurrence of every sampling point is a critical matter
and other algorithms might perform better.
In practical applications, a numerical experiment should be always conducted to choose a correct pattern generator and
to adjust variance value $\sigma^{2}$.

\begin{figure*}[h]
    \centering
    \includegraphics[scale=1.0]{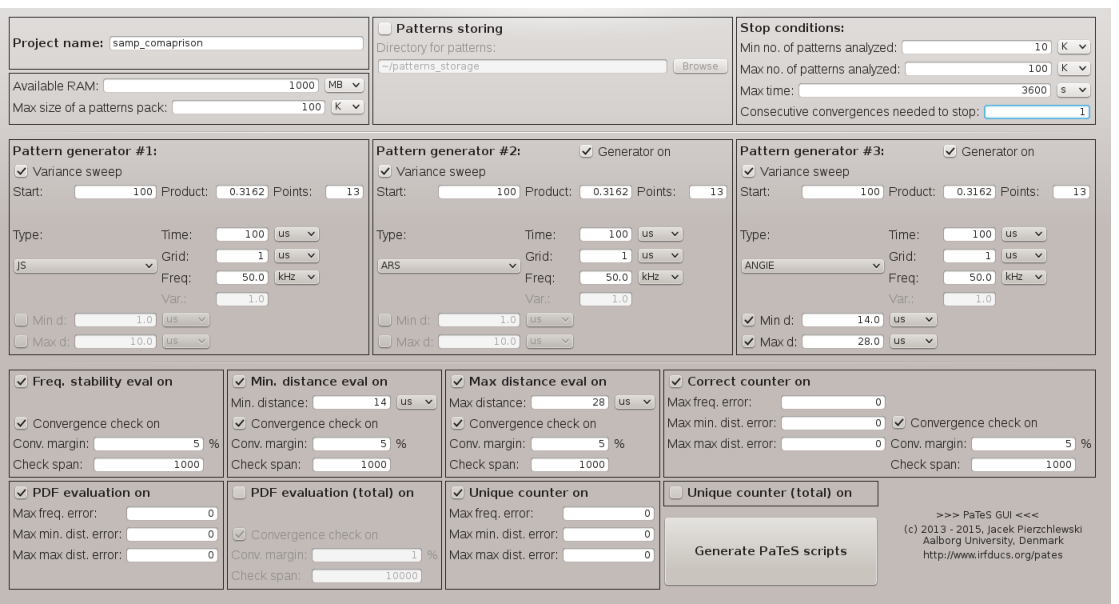}
    \caption{Graphical user interface to the Patterns Testing System (PATES).
             The system is available online in \cite{Jap01}.}
    \label{fig:pates-gui}
\end{figure*}

We prepared a software PAtterns TEsting System (PATES), which is open-source and available online \cite{Jap01}.
This software contains all the three generators considered in this paper plus routines which compute the proposed
quality parameters.
With this software a user is able to test the generators for his own sampling scenario.
We have created a graphical user interface to the software (Fig. \ref{fig:pates-gui}),
which makes using the system more intuitive.
Reproducible research scripts which can be used to produce results from the presented experiments
are also available in \cite{Jap01}.

%-------------------------------------------------------------------------
%  ----- ----- ----- IMPLEMENTATION ISSUES
%-------------------------------------------------------------------------

\section{Implementation issues}
\label{sec:implem}
In this Section we discuss some of the implementation issues of random sampling patterns.
In this paper, we focus on offline sampling pattern generation (Fig. \ref{fig:gen_case2}),
where patterns are prepared offline by a computational server and then stored in a
memory which is a part of a signal processing system.
Immediate generation of sampling patterns would require very fast pattern generators
which are able to generate every sampling point in a time much shorter than minimum time between sampling points $t_{\text{min}}$.
The ANGIE algorithm (Alg. \ref{alg:ANGIE}) requires a number of floating point computations before every sampling point is computed,
therefore very powerful computational circuit would be necessary in real time applications where $t_{\text{min}} < 1\mu s$.

% 6.1
\subsection{Software patterns generator}
In practical applications there is a need to generate $N \gg 1$ sampling patterns.
Sampling patterns are generated offline (Fig. \ref{fig:gen_case2})
on a computational server.
In naive implementation, Alg. \ref{alg:ANGIE}
is repeated $N$ times to generate $N$ random sampling patterns.
This approach is suboptimal,
because computation of initial parameters from equations (\ref{eqalg:precomp1}) and (\ref{eqalg:2ndPrep})
(lines 2-3) is unnecessarily repeated $N$ times.
In the optimal implementation
lines 2-3 are performed only once before a bag of patterns is generated.

\begin{figure*}[h]
    \centering
    \includegraphics[scale=1.0]{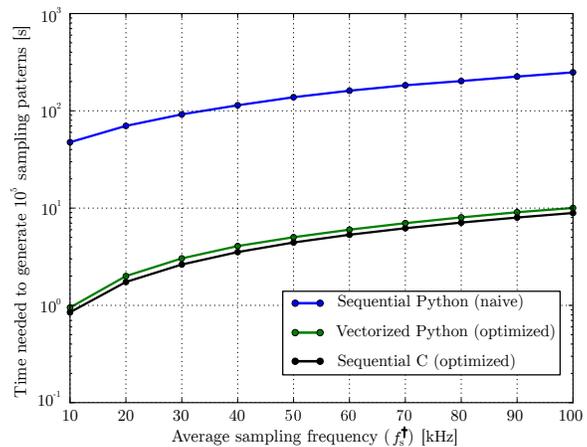}
    \caption{Time [seconds] needed to generate $10^5$ sampling patterns vs. the average sampling frequency of sampling patterns.}
    \label{fig:time_of_exec}
\end{figure*}

We implemented the ANGIE algorithm (naive implementation) in Python.
Furthermore, we prepared an implementation in C and an optimized implementation in Python (vectorized code).
All the implementations are available for download at \cite{Jap01}.
Fig. \ref{fig:time_of_exec} shows time needed to generate $N=10^5$
sampling patterns.
Parameters of sampling patterns are identical to the parameters
used in the experiment described in Section \ref{sec:expsetup}.
The average sampling frequency is swept from 10 $\kHz$ to 100 $\kHz$,
and the duration of the patterns is kept fixed.
Measurements were made on an Intel Core i5-3570K CPU, and a single core of the CPU was used.

The ANGIE algorithm operates mostly on integer numbers, and therefore
it requires maximally only three floating point operations p. sampling point.
The algorithm time complexity vs. the average sampling frequency of a pattern
is $O(n)$ (consider the logarithmic vertical scale),
because lines 5-13 in Alg. 2 are repeated
for every sampling point which must be generated.
As expected, the optimized vectorized Python / optimized C implementation
is much faster than the naive Python implementation.

% 6.2
\subsection{Driver of an analog-to-digital converter}
The analog-to-digital converter (ADC) driver is a digital circuit which
triggers the converter according to a given sampling pattern.
The maximum clock frequency of the driver determines the minimum grid period.
Detailed construction of the driver depends on the used ADC
because the driver must generate specific signals which drive the ADC.

\begin{figure*}[h]
    \centering
    \includegraphics[scale=1.0]{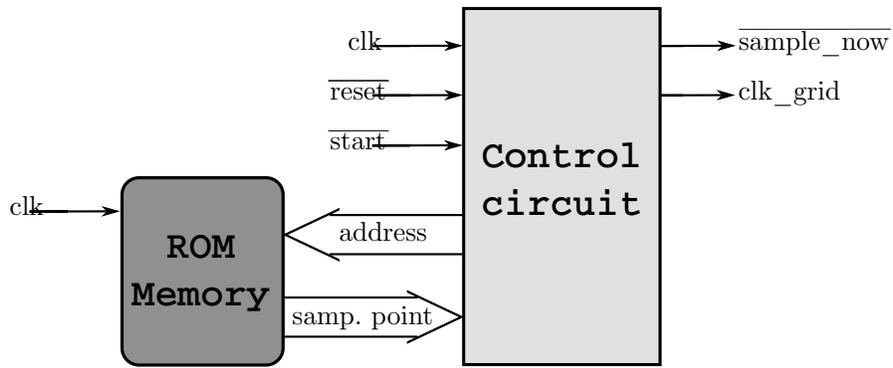}
    \caption{Block diagram of an implemented ADC driver.}
    \label{fig:driver}
\end{figure*}

A simple driver marks the 'sample now' signal every time the grid counter reaches
a value equal to the current sampling time point.
Such a driver was implemented in VHDL language.
The structure of the driver is shown in Fig. \ref{fig:driver}.
Due to the internal structure of the control circuit, the grid period is eight
times longer then the input clock period.
Table 2 contains results of synthesis of the driver
in four different Xilinx FPGAs.

        \begin{table}[h!]
        \centering
          \begin{tabular}{ c || c | c }
          Xilinx  & Max clock & Min grid  \\
            FPGA & frequency [MHz] & period $T_{\text{g}}$ [ns] \\
            \hline
            Spartan 3 & 439.97  & 18.2 \\
            Virtex 6  & 1078.98 & 7.4  \\
            Artix 7   & 944.47  & 8.5  \\
            Zynq 7020    & 1160.36 & 6.9  \\
            \hline
          \end{tabular}
          \caption{Maximum clock values and minimum grid periods of an implemented driver in different Xilinx FPGAs}
          \label{table:driver}
          \end{table}

\begin{figure*}[h]
    \centering
    \includegraphics[scale=1.0]{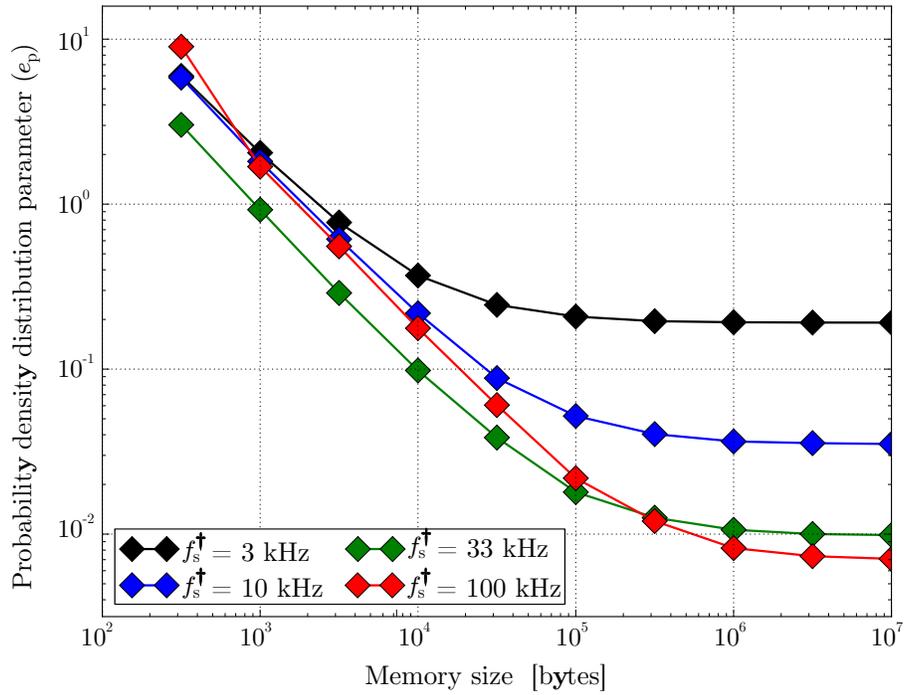}
    \caption{Probability density parameter $e_{\text{p}}$ (\ref{eq:PDFpar}) found for patterns generated by the ANGIE algorithm vs. the size of memory
    for patterns storing.}
    \label{fig:errPDF_vs_memory}
\end{figure*}

Sampling patterns are read from a ROM.
The amount of memory $n_{\text{m}}$ used to store a sampling pattern [in bytes] is:
\begin{equation}
n_{\text{m}} = K_{\text{s}} \cdot \left \lceil \frac{\log_2{ K_{\text{g}}}}{8} \right \rceil
\end{equation}
where $K_{\text{g}}$ is the number of grid points in a pattern and
$K_{\text{s}}$ is the number of sampling points in a pattern.
Depending on the available size of memory, different numbers of sampling patterns can be stored.
Fig. \ref{fig:errPDF_vs_memory} shows the relation between the memory size and the
probability density parameter $e_{\text{p}}$ (\ref{eq:PDFpar}) computed for
the proposed ANGIE algorithm.
The parameters of the sampling patterns are identical to the parameters used in
the experiment described in Section \ref{sec:expsetup},
although four different average sampling frequencies are used.

As expected, the higher the average sampling frequency of patterns,
the better the distribution of probability density function (parameter $e_{\text{p}}$ is lower).
The higher the average sampling frequency of patterns,
the more the memory needed to achieve the best possible
probability density parameter $e_{\text{p}}$.
If the available memory is low, the probability density function
becomes less equi-probable.

%-------------------------------------------------------------------------
%  ----- ----- ----- CONCLUSIONS
%-------------------------------------------------------------------------
\section{Conclusions}
\label{sec:conclusions}
This paper discussed generation of random sampling patterns
dedicated to event-driven ADCs.
Constraints and requirements for random sampling patterns and pattern generators were discussed.
Statistical parameters which evaluate sampling pattern generators were introduced.
We proposed a new algorithm which
generates constrained random sampling patterns.
The patterns generated by the proposed algorithm were compared with patterns generated by the state-of-the-art algorithms (Jittered Sampling and Additive Random Sampling).
It was shown, that the proposed algorithm performs better
in generation of random sampling patterns dedicated to event-driven ADCs.
Implementation issues of the proposed method were discussed.

%-------------------------------------------------------------------------
%  ----- ----- ----- ACKNOWLEDGEMENT
%-------------------------------------------------------------------------
\section*{Acknowledgment}
The work is supported by The Danish Council for Independent Research under
grant number 0602--02565B.

%-------------------------------------------------------------------------
%  ----- ----- ----- FIGURES
%-------------------------------------------------------------------------
\newpage

\newpage
\begin{figure*}[h]
    \includegraphics[scale=1.0]{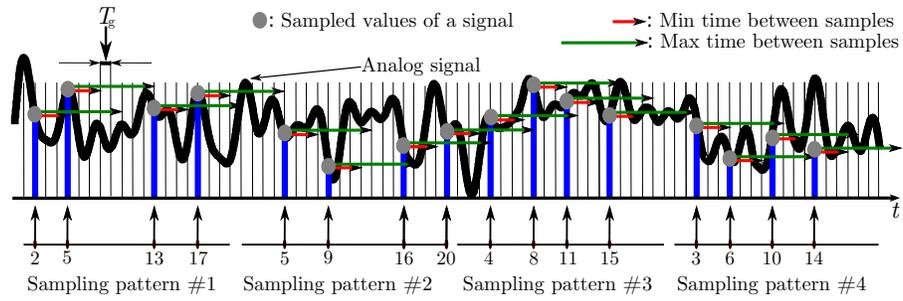}
    \caption{Example of constrained random sampling patterns applied to an analog signal.
           There is a minimum (red arrow) and maximum (green arrow) allowed interval between sampling points.
            Furthermore, every pattern has the equal number of sampling points.}
    \label{fig:pattern_constrained}
\end{figure*}

\newpage
\begin{figure*}[h]
    \centering
    \includegraphics[scale=1.0]{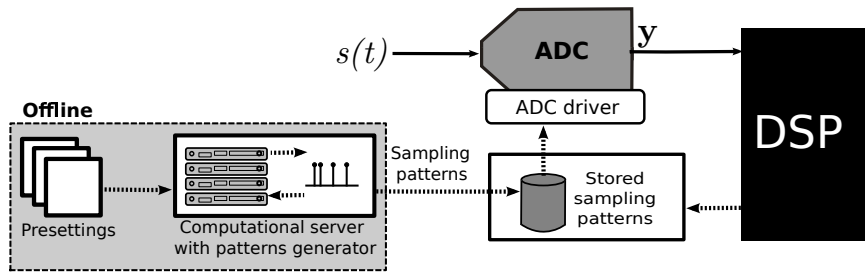}
   \caption{Offline generation of sampling patterns. Sampling patterns are prepared offline on a computational server, and then stored in a memory in the sampling system.}
    \label{fig:gen_case2}
\end{figure*}

\newpage

\newpage
\begin{figure*}[h]
   \centering
    \includegraphics[scale=1.0]{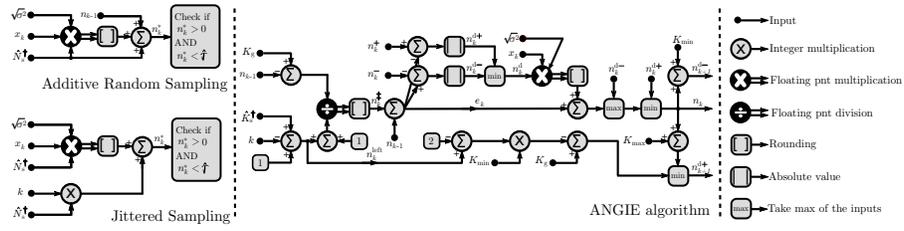}
    \caption{Block diagram showing the generation of one sampling point in the Additive Random Sampling, the Jittered Sampling and the ANGIE algorithm.}
    \label{fig:gen_case2}
\end{figure*}

\newpage

\newpage

\newpage

\newpage

\newpage

\newpage

\newpage

\newpage

\newpage

\newpage

\newpage

\newpage

\end{document}